\documentclass[%
 preprint,
superscriptaddress,
 amsmath,amssymb,
 aps,
]{revtex4-2}

\usepackage{graphicx}
\usepackage{dcolumn}
\usepackage{bm}
\usepackage{xcolor}
\usepackage{subcaption}

\begin{document}


\title{Structural Origin of Water’s Heat Capacity Anomaly from Classical and Quantum Simulations}%

\author{Kam-Tung Chan}
\affiliation{Department of Chemistry, University of California, Davis, Davis, CA, 95616, USA}

\author{Dylan A. Folkner}
\affiliation{Department of Chemistry, University of California, Davis, Davis, CA, 95616, USA}

\author{Zekun Chen}
\affiliation{Energy Storage and Distributed Resources Division, Lawrence Berkeley National Laboratory, Berkeley, California 94720, USA}

\author{Margaret L. Berrens}
 \affiliation{Quantum Simulations Group, Physics Division, Lawrence Livermore National Laboratory, Livermore, California 94550, USA}
 
\author{Alexei A. Stuchebrukhov}
\affiliation{Department of Chemistry, University of California, Davis, Davis, CA, 95616, USA}

\author{Lee-Ping Wang}
\affiliation{Department of Chemistry, University of California, Davis, Davis, CA, 95616, USA}

\author{Davide Donadio}
 \email{Corresponding author: ddonadio@ucdavis.edu}
\affiliation{Department of Chemistry, University of California, Davis, Davis, CA, 95616, USA}

\date{\today}

\begin{abstract}
Water’s isobaric heat capacity is anomalously large under ambient conditions and exhibits a sharp maximum upon supercooling. Using classical and path-integral molecular dynamics with accurate machine-learning interatomic potentials, we show that nuclear quantum effects primarily act by suppressing high-frequency vibrations, while the anomalous temperature dependence of the isobaric heat capacity originates from structural fluctuations, quantified by the second-solvent-shell intruder order parameter. A simple two-state mapping reveals an effective enthalpy scale of $\approx$ 3$-$4 kJ mol$^{-1}$ associated with the interconversion of low- and high-density-like local structures, providing a microscopic link between their population changes and the excess heat capacity from supercooled to ambient conditions.
\end{abstract}

\maketitle



Heat capacity is one of many anomalous thermodynamic properties of liquid water and is central to its role in climate, geophysics, and biology~\cite{vallis_climate_2011, gallo_water_2016, Bagchi_2013}. In particular, the isobaric heat capacity ($C_{P}$) is anomalously large under ambient conditions.
For comparison, the $C_{P}$ of water at ambient conditions is 4.18 J~g$^{-1}$ K$^{-1}$, substantially larger than that of several common molecular liquids, such as methanol (2.56 J~g$^{-1}$K$^{-1}$), ethanol (2.52 J g$^{-1}$K$^{-1}$), $n$-hexane (2.29 J~g$^{-1}$K$^{-1}$), acetone (2.16 J g$^{-1}$K$^{-1}$), and aniline (2.06 J g$^{-1}$K$^{-1}$) \cite{naef_calculation_2019, nist_condensed_heat_capacity}.
$C_P$ also exhibits a pronounced maximum upon supercooling to $\sim$230 K at 1 bar \cite{poole_phase_1992, xu_relation_2005, pathak_enhancement_2021}. 
This behavior is among the most prominent thermodynamic anomalies of water and is closely tied to the unusual response of its hydrogen-bond (H-bond) network \cite{kim_maxima_2017, pathak_temperature_2019, yun_correlated_2022, muthachikavil_unraveling_2024}. 
Although earlier molecular simulations have qualitatively reproduced this behavior, a quantitative microscopic explanation connecting local structural fluctuations to both the supercooled maximum and the unusually high $C_{P}$ at ambient conditions remains elusive \cite{kim_maxima_2017, pathak_temperature_2019, eltareb_nuclear_2021, eltareb_evidence_2022, chen_thermodynamics_2024, muthachikavil_unraveling_2024, eltareb_isotope-substitution_2025, savoia_influence_2025}.

Classical molecular dynamics (MD) simulations of water generally overestimate $C_{P}$ because all vibrational modes are thermally populated~\cite{horn_development_2004, vega_heat_2010, reddy_accuracy_2016, ceriotti_nuclear_2016, eltareb_nuclear_2021, eltareb_evidence_2022, chen_thermodynamics_2024, eltareb_isotope-substitution_2025, savoia_influence_2025, xu_nep-mb-pol_2025}. 
In contrast, path-integral molecular dynamics (PIMD) simulations suppress the contribution of high-frequency modes, bringing the predicted heat capacity into closer agreement with experiments~\cite{vega_heat_2010, ceriotti_nuclear_2016, eltareb_nuclear_2021, eltareb_evidence_2022, eltareb_isotope-substitution_2025, savoia_influence_2025, xu_nep-mb-pol_2025}. These studies clarify the role of nuclear quantum effects (NQEs) in setting the absolute magnitude of $C_{P}$, but they do not, by themselves, explain why water has such a large heat capacity or how the anomalies under ambient and supercooled conditions are connected.

The liquid-liquid phase transition (LLPT) scenario offers a compelling interpretation of water's anomalies in terms of interconversion between low-density-like (LDL-like) and high-density-like (HDL-like) local structures~\cite{chau_new_1998, errington_relationship_2001, debenedetti_supercooled_2003, holten_entropy-driven_2012, russo_understanding_2014, pathak_temperature_2019, shi_anomalies_2020, muthachikavil_unraveling_2024, song_understanding_2026}.
In this picture, thermodynamic anomalies arise from enhanced fluctuations in local structure, but establishing a quantitative connection between a structural order parameter and the temperature dependence of $C_{P}$ remains an open challenge.

Here, we combine classical and path-integral molecular dynamics (PIMD) in the isothermal-isobaric (NPT) ensemble with accurate machine-learning interatomic potentials and show that the vibrational and structural contributions to water's $C_{P}$ separate naturally. 
NQEs primarily lower the overall magnitude of $C_{P}$ by suppressing high-frequency vibrational contributions, whereas its anomalous temperature dependence is captured by a structural order parameter, the ``solvation-shell averaged second solvation shell intruders" (SSSI$^{(n)}$), where $n$ refers to the number of shells on which the number of intruders is averaged (see section S1).
Choosing $n=3$, the SSSI$^{(3)}$ order parameter correlates well with water's mobility and distinguishes between LDL-like and HDL-like structures, successfully describing the LLPT in supercooled water~\cite{wang_direct_2026}.
By mapping SSSI$^{(3)}$ onto a simple two-state model, we extract an effective enthalpy scale of about 3$-$4 kJ mol$^{-1}$ between LDL-like and HDL-like local structures and decompose $C_{P}$ into a physically constrained baseline and an SSSI-based structural contribution that accounts for both the supercooled maximum, corresponding to crossing the Widom line, and the excess heat capacity near ambient temperature~\cite{widom_topics_1963,franzese_widom_2007,abascal_widom_2010}. This provides a microscopic framework, consistent with the LLPT scenario, for linking local structural topology to the thermodynamic anomalies of water.

Two neuroevolution potentials (NEPs) \cite{fan_neuroevolution_2021, fan_gpumd_2022} were trained separately on a van der Waals-corrected hybrid density functional (revPBE0-D3) \cite{perdew_generalized_1996, zhang_comment_1998, adamo_toward_1999, grimme_consistent_2010} and on an ab initio many-body potential energy function at the level of coupled-cluster with singles, doubles, and perturbative triples (MB-pol) \cite{babin_development_2013, babin_development_2014, medders_development_2014}. Training and computational details are reported in sections S2 and S3 of the Supplemental Material \cite{supplemental} (see also refs. \cite{schran_committee_2020, abascal_potential_2005, chen_thermodynamics_2024, berrens_nuclear_2024, kuhne_cp2k_2020, marsalek_quantum_2017, vandevondele_gaussian_2007, guidon_auxiliary_2010, goedecker_separable_1996, riera_mbx_2023, morawietz2016van, omranpour2024perspective, bussi_canonical_2007, bernetti_pressure_2020, bussi_accurate_2007, berendsen_molecular_1984, ceriotti_efficient_2010, luzar_hydrogen-bond_1996} therein). Hereafter, we present results from the model trained on MB-pol (NEP3@MB-pol), while results from the model trained on revPBE0-D3 are shown in section S9 of the Supplemental Material~\cite{supplemental}.


\begin{figure}[t]
\includegraphics[width=0.48\textwidth]{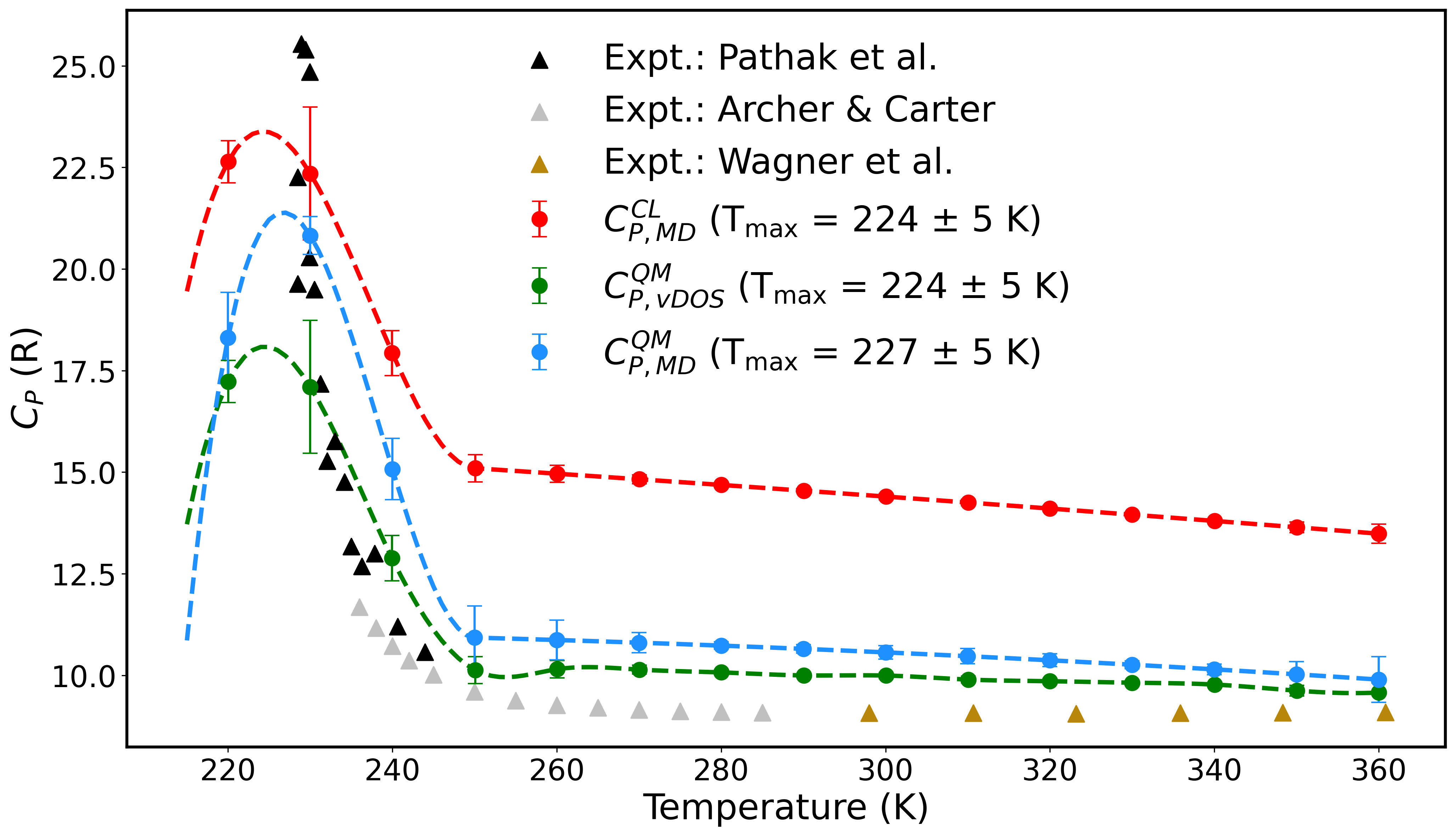}
\caption{\label{Cp_mb-pol} $C_{P}$ calculated directly using Equation \ref{cp_def} from polynomial fits to classical MD {(red; $C^{CL}_{P, MD}$) and PIMD (blue; $C^{QM}_{P, MD}$)} simulations with the NEP3@MB-pol potential. The $C_{P}$ obtained with the quantum correction of the $vDOS$ in classical MD (green; $C^{QM}_{P,vDOS}$) is also shown. Error bars indicate one standard deviation, estimated by propagating the standard deviation of the block-averaged enthalpy through Equation \ref{cp_def}. The uncertainty of $\pm$ 5 K in T$_{\mathrm{max}}$ reflects the half-spacing of the temperature grid (10 K). Experimental values are shown as triangles \cite{pathak_enhancement_2021, archer_thermodynamic_2000, wagner_iapws_2002}.
}
\end{figure}
We calculated the isobaric heat capacity from its definition 
\begin{eqnarray}
C_{P}(T)=\left(\frac{\partial H}{\partial T}\right)_{P}, \label{cp_def}
\end{eqnarray}
where the enthalpy $H(T)$ was estimated directly from classical MD ($CL$) and PIMD ($QM$) simulations, then fitted to a piecewise function with two sets of polynomials: a fourth-order polynomial for $T<250$ K and a third-order polynomial for $T\geq250$ K (see section S4)~\cite{chen_thermodynamics_2024}.
As shown in Figure~\ref{Cp_mb-pol}, classical MD simulations overestimate $C_{P}$ by 4$-$5 R at ambient temperature relative to experiments \cite{archer_thermodynamic_2000, wagner_iapws_2002}. 
PIMD simulations predict lower $C_{P}$ values that are in closer agreement with experiments, consistent with previous simulations~\cite{xu_nep-mb-pol_2025}. 

The impact of NQEs on $C_{P}$ can also be estimated using a quantum correction scheme based on the vibrational density of states (vDOS) obtained from classical MD simulations, by integrating the power spectrum of the mass-weighted velocity autocorrelation function ($S(\omega)$) and reweighting it by the difference between classical and quantum occupations~\cite{berens_thermodynamics_1983, savoia_influence_2025}:
\begin{equation}
    C^{QM}_{P,vDOS} = C^{CL}_{P, MD} - \Delta C_{P, vDOS},
\label{cp_corr}
\end{equation}
where the correction is: 
\begin{equation}
\Delta C_{P, vDOS} = \int^{\infty}_{0} k_{B} S(\omega)\left[1- \frac{(\hbar\omega/k_{B}T)^{2}e^{\hbar\omega/k_{B}T}}{(e^{\hbar\omega/k_{B}T}-1)^{2}}\right]d\omega.
\label{delta_cp}
\end{equation}
This vDOS-corrected $C_{P}$ ($C^{QM}_{P,vDOS}$) is lower than the classical $C_{P}$ obtained directly from the polynomial fit ($C^{CL}_{P, MD}$) and brings it closer to experiment by 4$-$5 R under ambient conditions, consistent with the PIMD results (see Figure \ref{Cp_mb-pol}). This shows that including NQEs via PIMD brings the $C_{P}$ curve substantially closer to experiment across the full temperature range, largely through suppression of high-frequency vibrational contributions {(see section S5)}. 
However, both classical MD and PIMD simulations reproduce a pronounced maximum in $C_{P}$ near 230 K in the supercooled regime, in good agreement with experiments~\cite{archer_thermodynamic_2000, pathak_enhancement_2021}. 
This suggests that NQEs have a negligible effect on the temperature of maximum $C_{P}$ (T$_{max}$), though the temperature uncertainty of $\pm 5$~K should be kept in mind. Pinning down the value of the $C_{P}$ maximum would require a finer temperature grid, but the current estimate is compatible with experiments on supercooled water in confinement~\cite{pathak_enhancement_2021}.

\begin{figure}[b]
    \includegraphics[width=0.48\textwidth]{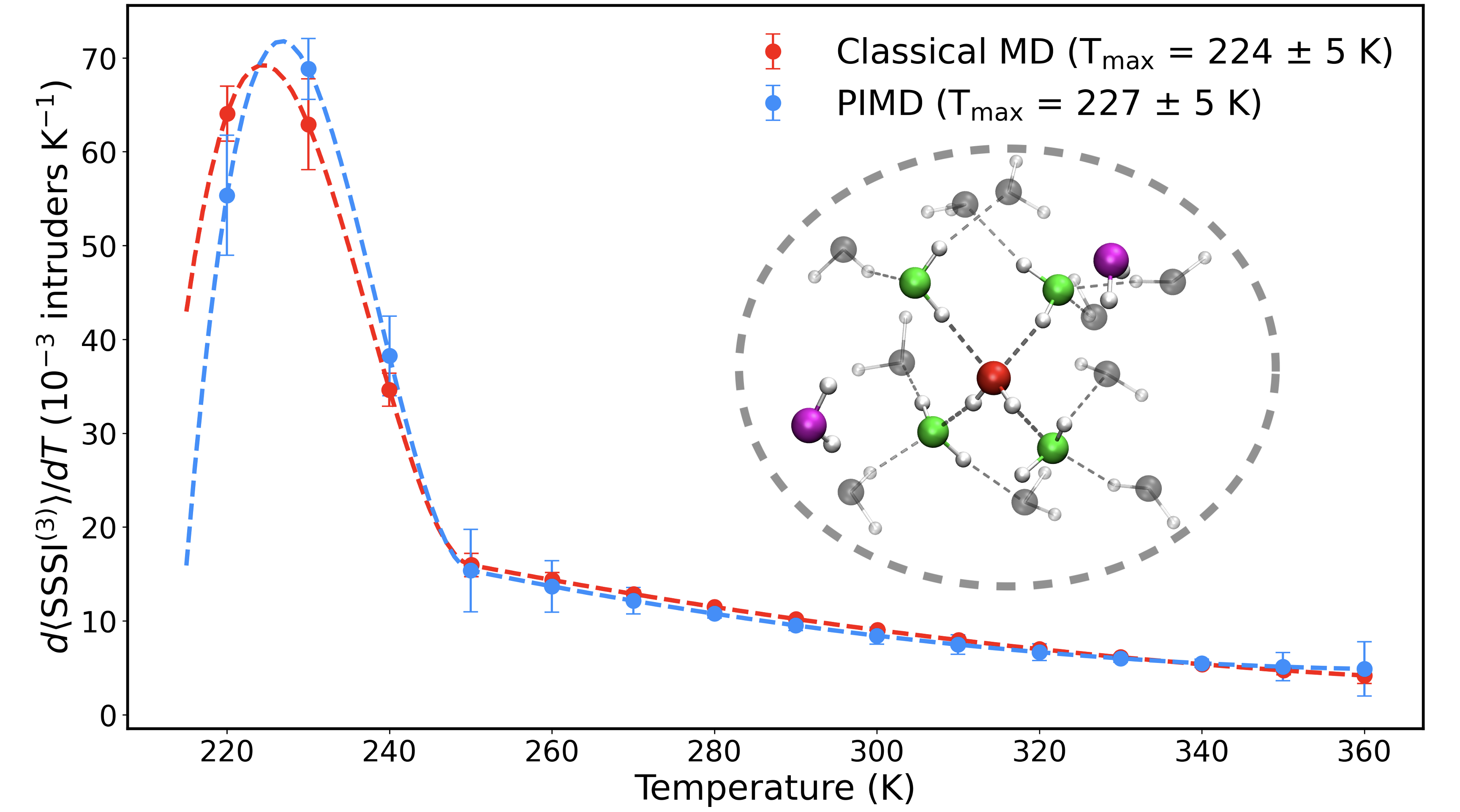}
    \vfill
    \includegraphics[width=0.48\textwidth]{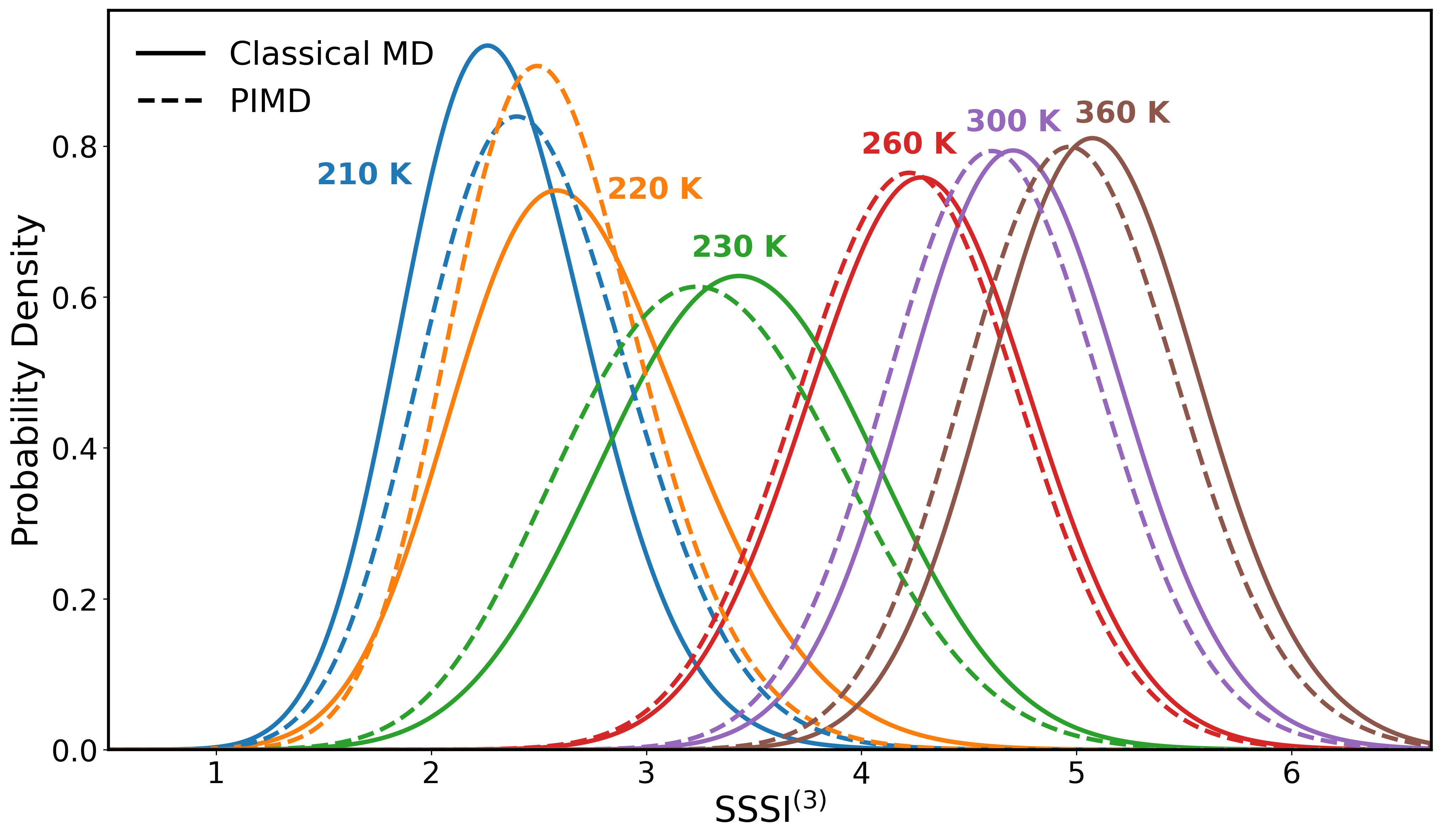}
    \caption{
    (Top panel) $\frac{d\langle\mathrm{SSSI^{(3)}}\rangle}{dT}$ obtained from classical MD (red) and PIMD (blue) simulations with the NEP3@MB-pol potential. Error bars correspond to one standard deviation estimated from the block-averaged SSSI$^{(3)}$. The uncertainty of $\pm$ 5 K in T$_{\mathrm{max}}$ reflects the half-spacing of the simulated temperature grid, sampled every 10 K. 
    {The inset shows a representative snapshot of a central water molecule (red) with two SSSI (magenta) in its solvation shell (green: first solvation shell; grey: second solvation shell) from the classical MD simulation at 210 K. H-bonds are shown as dashed lines.} 
    (Bottom panel) Distributions of SSSI$^{(3)}$ at different temperatures from classical MD (solid lines) and PIMD (dashed lines) simulations. 
    }
    \label{SSSI3_results_mb-pol}
\end{figure}
Next, we calculated the SSSI$^{(3)}$ order parameter as a function of temperature to investigate how the topology of the molecular network is connected to the macroscopic $C_{P}$. 
{SSSI is defined as the number of water molecules in the second solvation shell not contained in the union of first solvation shells of water molecules in the first solvation shell, where the first and solvation shells are defined as the closest 4 and 16 water molecules respectively (see the inset in Figure \ref{SSSI3_results_mb-pol} (top panel) and section S1). SSSI$^{(3)}$ is then computed by averaging the SSSI over molecules in the first solvation shell three times.}
Figure~\ref{SSSI3_results_mb-pol} (bottom panel) shows the SSSI$^{(3)}$ probability distributions at different temperatures. 
Low SSSI$^{(3)}$ values correspond to LDL-like local environments, where fewer water molecules penetrate the second-shell region, whereas large SSSI$^{(3)}$ values indicate HDL-like environments with more second-shell intruders. As temperature increases, the SSSI$^{(3)}$ distribution shifts continuously toward higher values, reflecting the gradual conversion from LDL-like to HDL-like local structures. The differences between the classical MD and PIMD distributions are small across all temperatures, consistent with the expectation of moderate NQEs on the local structure of water \cite{stolte_nuclear_2024, chiang_experimental_2025}. 
Except in the deepest supercooling case (210~K), the PIMD distributions are centered at slightly lower SSSI$^{(3)}$ average values, indicating that NQEs mildly suppress local structural disorder.  
Near 230 K, where $C_{P}$ reaches its maximum, the distribution is broad and centered around SSSI$^{(3)}$ $\approx$ 3.3$-$3.4, suggesting that SSSI$^{(3)}\approx 3$ marks the crossover between LDL-like and HDL-like environments.

To verify the connection between the $C_{P}$ maximum and the change in SSSI$^{(3)}$, we obtained the derivative of the mean SSSI$^{(3)}$ ($\langle\mathrm{SSSI^{(3)}}\rangle$) with respect to temperature ($\frac{d\langle\mathrm{SSSI^{(3)}}\rangle}{dT}$) using the same method as for $C_{P}$, by fitting $\langle\mathrm{SSSI^{(3)}}\rangle$ as a piecewise function of temperature (see section S4). 
The resulting curve closely tracks the temperature-dependent $C_{P}$, and its low-temperature peak coincides with the $C_{P}$ maximum in both classical MD and PIMD simulations (see Figure \ref{SSSI3_results_mb-pol} (top panel)). 
This correlation indicates that changes in local structure, quantified by the $\mathrm{SSSI^{(3)}}$ order parameter, are closely associated with the anomalous temperature dependence of $C_{P}$.  
The corresponding value of $\langle\mathrm{SSSI^{(3)}}\rangle$ at the maximum is $\sim$ 3 in both classical MD and PIMD simulations, consistent with the transition value between the LDL and HDL phases near the critical point determined in the previous study \cite{wang_direct_2026}. 
{Consistently, the variance of SSSI$^{(3)}$ and the enthalpy–SSSI$^{(3)}$ correlation are also strongly enhanced in this temperature range (see section S6).}
{Together, these observations support the interpretation that the peak in $\frac{d\langle\mathrm{SSSI^{(3)}}\rangle}{dT}$, and therefore the maximum in $C_{P}$, arises from the most rapid change in the population of these two local structural motifs.}
As the temperature increases beyond $\sim$ 230 K, the magnitude of the derivative decreases but remains positive, 
{suggesting that interconversion between LDL-like and HDL-like local structures still contributes to the high $C_{P}$ at high temperature.}

\begin{figure}[tb]
\includegraphics[width=0.48\textwidth]{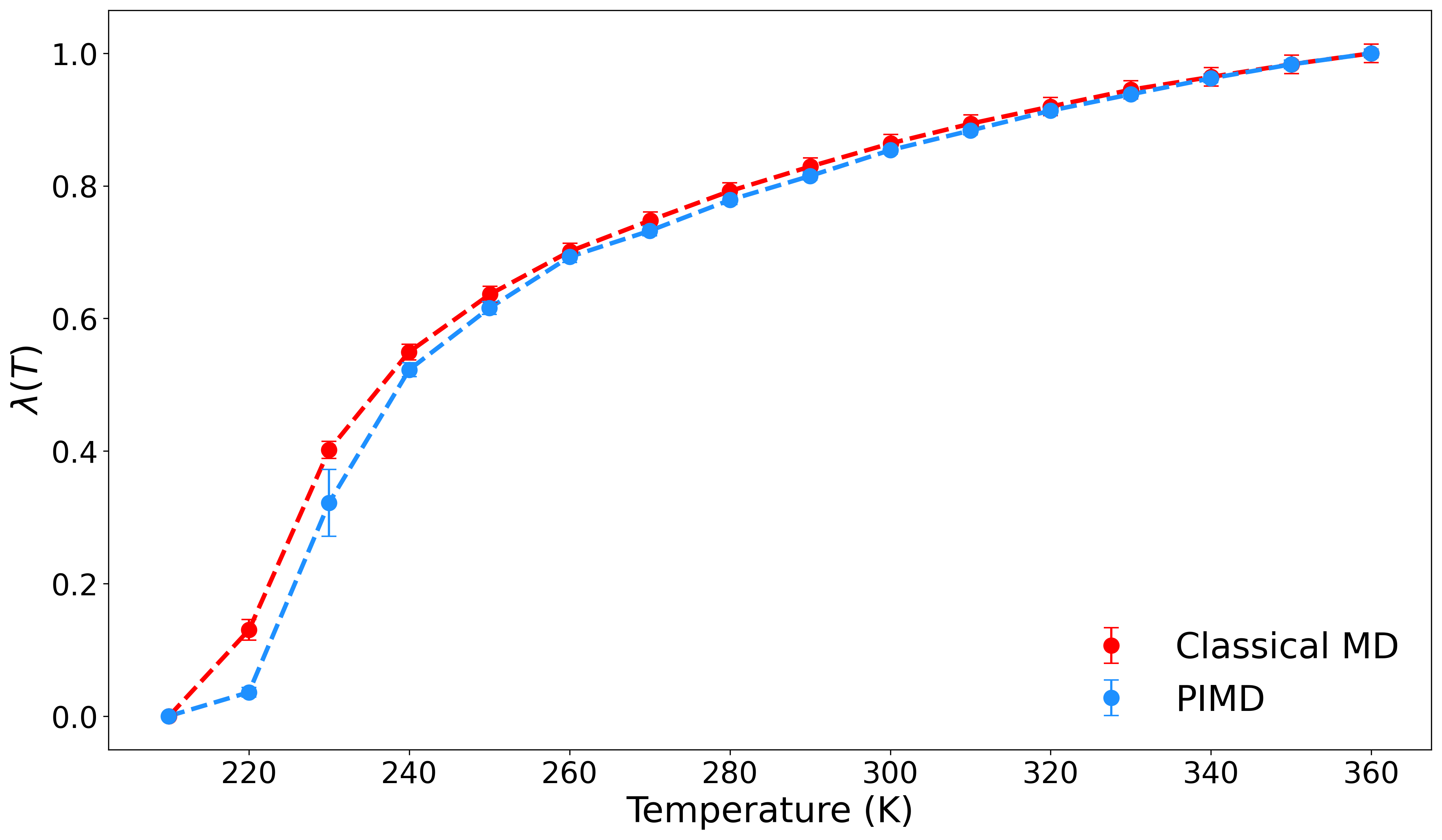}
\caption{Fraction of HDL-like water ($\lambda(T)$) obtained using Equation \ref{HDLL_fraction} from classical MD (red) and PIMD (blue) simulations with the NEP3@MB-pol potential. Error bars indicate one standard deviation, estimated by propagating the standard deviation of the block-averaged SSSI$^{(3)}$ through Equation \ref{HDLL_fraction}.} 
\label{HDL-like_fraction_mb-pol}
\end{figure}

To further validate our interpretation of $Cp$ in terms of a local structural transition from low- to high-density liquid, we use a simple two-state model, assuming that the average value of the order parameter can be written as:
\begin{equation}
\langle\mathrm{SSSI^{(3)}}\rangle(T) = \lambda(T)\mathrm{SSSI^{(3)}_{HDLL}} + [1-\lambda(T)]\mathrm{SSSI^{(3)}_{LDLL}},
\label{HDLL_fraction}
\end{equation}
where $\lambda(T)$ is the fraction of HDL-like local structures. This construction provides an effective two-state representation of the simulated structural evolution. 
$\mathrm{SSSI^{(3)}_{LDLL}}$ and $\mathrm{SSSI^{(3)}_{HDLL}}$ are defined as $\langle\mathrm{SSSI^{(3)}}\rangle(T=210 \mathrm{K})$ and $\langle\mathrm{SSSI^{(3)}}\rangle(T=360 \mathrm{K})$, respectively.
As shown in Figure \ref{HDL-like_fraction_mb-pol}, $\lambda(T)$ increases most sharply at the transition temperature, consistent with the distributions in Figure \ref{SSSI3_results_mb-pol} (bottom panel), and reflecting the most rapid crossover between LDL-like and HDL-like environments.

\begin{figure*}
\subcaptionbox{Classical MD\label{fig:Cp_Decomp_CL}}{\includegraphics[width=0.495\textwidth]{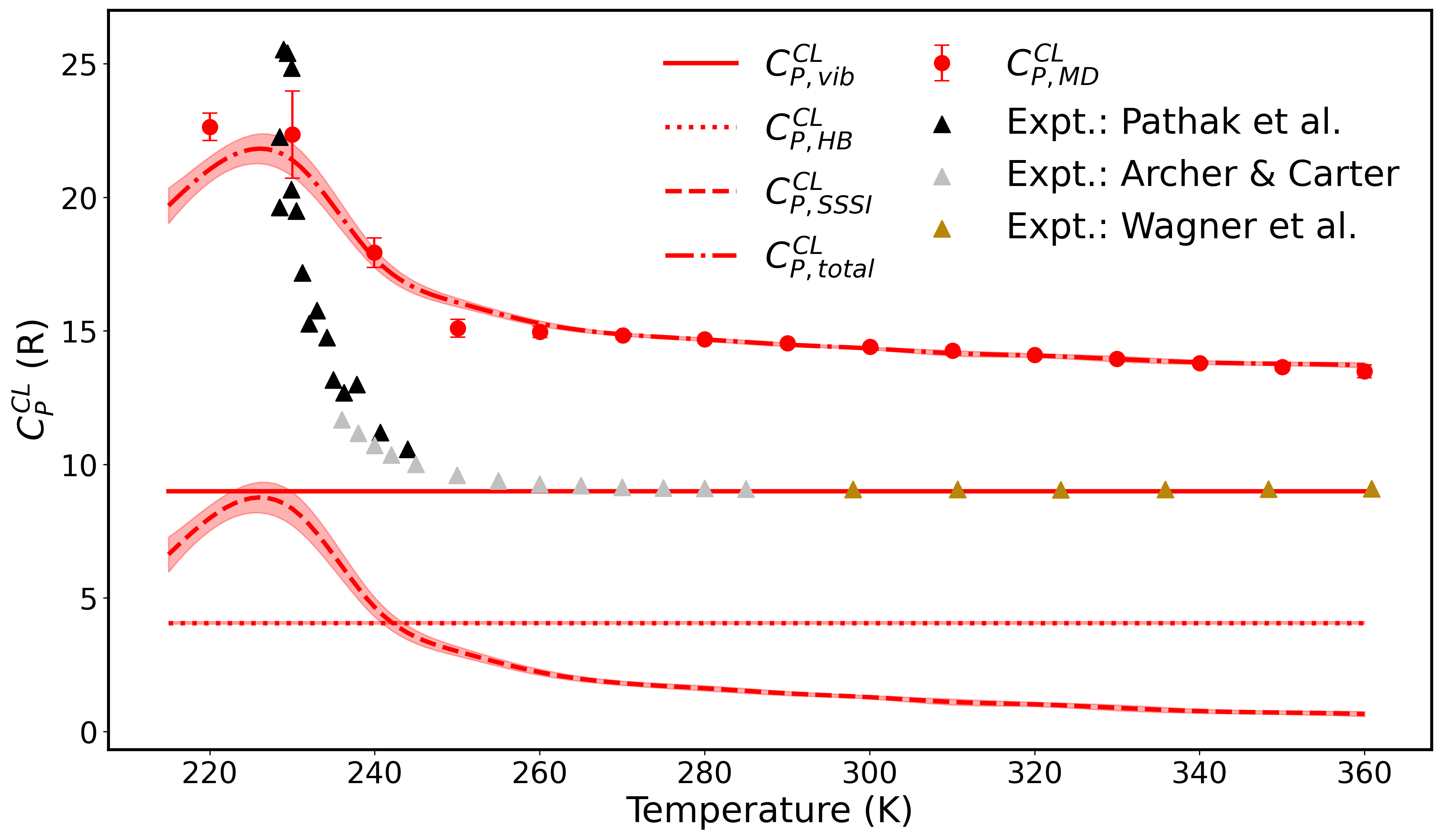}}
\hfill
\subcaptionbox{PIMD\label{fig:Cp_Decomp_QM}}{\includegraphics[width=0.495\textwidth]{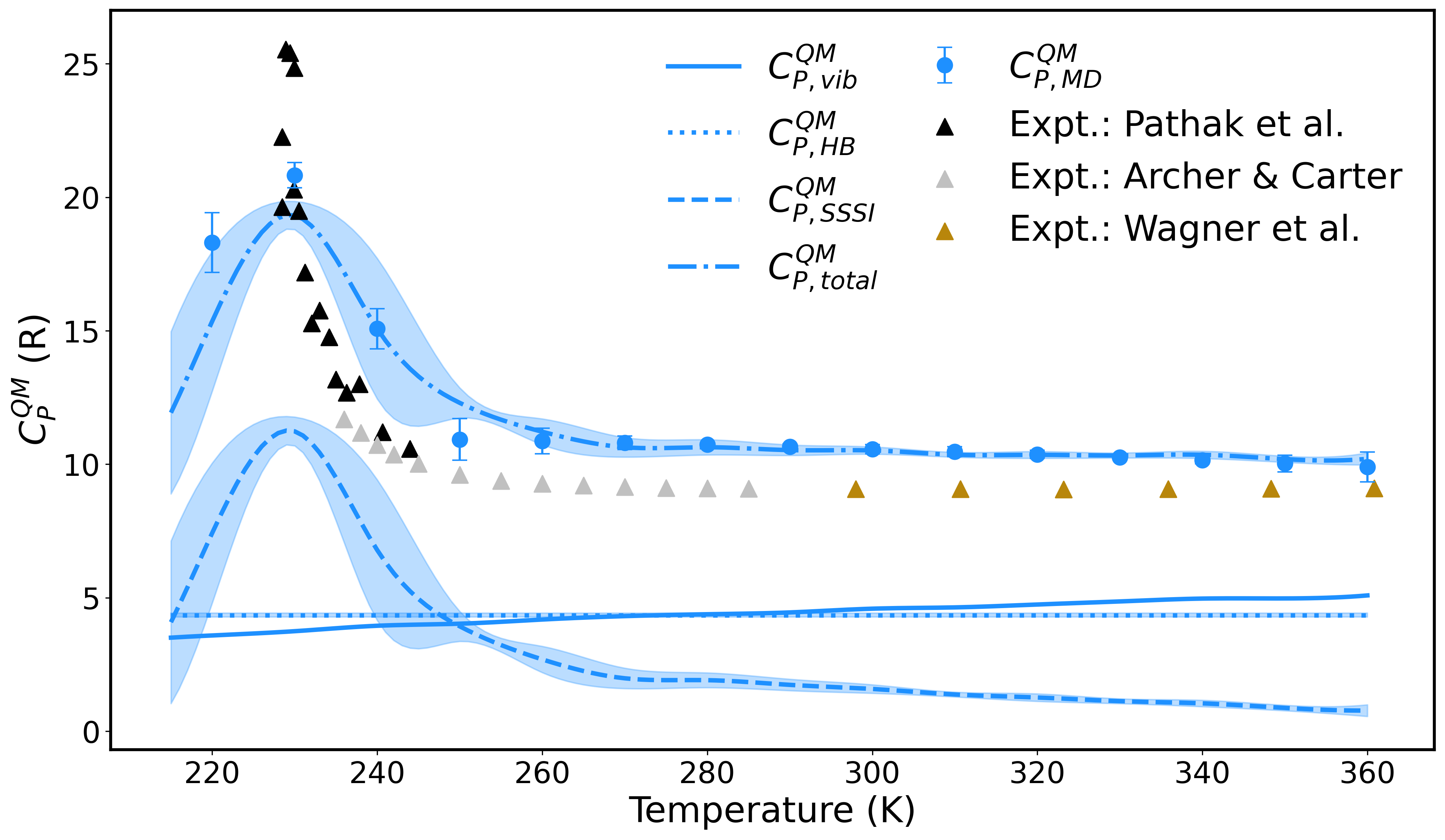}}
\caption{{$C_{P}$ from (a) classical MD ($C_{P}^{CL}$) and (b) PIMD ($C_{P}^{QM}$) simulations with the NEP3@MB-pol potential, calculated using Equation \ref{cp_def} by decomposing $H(T)$ into $H_{base}$ and $H_{SSSI}$. Errors indicate one standard deviation, estimated by propagating the standard deviations of the block-averaged enthalpy and SSSI$^{(3)}$ through Equation \ref{cp_def}. The $C_{P}$ from Figure \ref{Cp_mb-pol}, calculated by fitting $H(T)$ with the piecewise function, is also shown ($C_{P, MD}$). Experimental values are shown as triangles \cite{pathak_enhancement_2021, archer_thermodynamic_2000, wagner_iapws_2002}.}}
\label{Cp_Decomp_mb-pol}
\end{figure*}

Finally, we demonstrate that the effective two-state model can be used to retrieve the temperature dependence of $C_{P}$.
To this end, we decompose $H(T)$ into a physically constrained baseline ($H_{base}(T)$) and an SSSI-based structural contribution arising from the interconversion between LDL-like and HDL-like local structures ($H_{SSSI}(T) = \Delta H_{L/H}\cdot \lambda(T)$):
\begin{eqnarray}
    &&H(T) = H_{base}(T) + H_{SSSI}(T) \label{H_decomp} \\
    &&= H_0 + \int_{T_{ref}}^{T}C_{P, base}(T')dT'+ \Delta H_{L/H}\cdot \lambda(T). \nonumber
\end{eqnarray}
where $H_0$ is the baseline enthalpy offset at a reference temperature ($T_{ref}$), and: 
\begin{equation}
C_{P, base}(T) = C_{P, vib}(T) + C_{P, HB}(T). \label{Cp_base}
\end{equation}
Here, $C_{P, vib}$ is the vibrational contribution to $C_{P}$ from the vDOS, which is exactly 9R in the classical case, and $C_{P, HB}$ is a residual background contribution to $C_{P}$, which encompasses configurational and anharmonic effects. 
{This term ($C_{P,HB}$) is mostly related to the enthalpy cost of breaking H-bonds. The average number of H-bonds per molecule ($\langle N_{HB}\rangle$) decreases linearly with $T$ for $T\geq$ 250~K (Figure S5 and Table S4), so we can roughly assume that C$_{P, HB}\sim H_{HB} |d\langle N_{HB}\rangle/dT|$ is temperature independent (see section S7 for the detailed analysis). The fitted values of  C$_{P, HB}$ are 4.1 and 4.3 in classical and quantum simulations, which, with a calculated $|d\langle N_{HB}\rangle/dT| \approx 0.0026 \times 10^{-3}$ H-bonds K$^{-1}$, give an estimate of the average H-bonding enthalpy, $H_{HB} = 0.14$~eV (Table S4), compatible with literature values~\cite{reddy_accuracy_2016}.}
$\Delta H_{L/H}$ is the enthalpy of transition from 
LDL-like to HDL-like local structures as defined by the SSSI order parameter.
$H(T)$ was fitted to Equation~\ref{H_decomp} over the entire simulated temperature range, and $C_P$ was determined through Equation \ref{cp_def}. 
As shown in Figure \ref{Cp_Decomp_mb-pol}, the total quantum $C_P$ ($C_{P, total}^{QM}$) at ambient temperatures is closer to experiment than the total classical $C_P$ ($C_{P, total}^{CL}$), mainly through suppression of the vDOS contribution ($C_{P, vib}$), consistent with the findings in Figure \ref{Cp_mb-pol}. $C_{P, total}$ is also consistent with $C_{P, MD}$ obtained directly from the classical MD and PIMD simulations shown in Figure \ref{Cp_mb-pol}.
We also observe that the shape of the SSSI-based structural contribution ($C_{P, SSSI}$) reproduces the maximum at $\sim$ 230 K. This behavior supports the interpretation that the supercooled heat capacity anomaly is dominated by LDL-like/HDL-like interconversion within the present two-state mapping.
In addition, the trend of $C_{P, SSSI}$ is consistent with the enthalpy–SSSI$^{(3)}$ correlation (see section S6), indicating that SSSI$^{(3)}$ does not account for the full heat capacity over the entire temperature range. Rather, it captures the structurally correlated component of the enthalpy fluctuations associated with LDL-like/HDL-like interconversion.

Depending on how $\mathrm{SSSI^{(3)}_{LDLL}}$ and $\mathrm{SSSI^{(3)}_{HDLL}}$ are defined, $\Delta H_{L/H}$ varies slightly, from $2.8$ to $3.3$ kJ mol$^{-1}$ for classical MD simulations and from $3.6$ to $3.9$ kJ mol$^{-1}$ for PIMD simulations (see Tables S5 and S6). 
$\Delta H_{L/H}$ should therefore be regarded only as an effective scaling difference between the LDL-like and HDL-like local structures, and the present two-state model should be interpreted as an effective representation of a continuous structural evolution rather than a sharp phase transition.
Nevertheless, $\Delta H_{L/H}$ is physically reasonable: it is larger than, but of the same order of magnitude as, the energy difference between low-density amorphous (LDA) and high-density amorphous (HDA) ice computed with our NEP3@MB-pol potential at 100 K ($1.7\pm 0.2$ kJ mol$^{-1}$), and broadly consistent with the energy difference between high-density and glassy states recently computed by Pabst and Hassanali ($\sim$ 1–2 kJ mol$^{-1}$) \cite{pabst2026glassy}. 
{The larger value of $\Delta H_{L/H}$ may be related to the local nature of the SSSI$^{(3)}$ order parameter and to the specific choice of averaging over three solvation shells.}

In summary, classical MD and PIMD simulations with accurate machine-learning interatomic potentials show that NQEs primarily reduce the absolute magnitude of $C_{P}$ by suppressing high-frequency vibrational contributions, while leaving the temperature of the $C_{P}$ maximum nearly unchanged. In contrast, the anomalous temperature dependence of $C_{P}$, from the supercooled peak to its unusually large value near ambient conditions, is closely tracked by the temperature derivative of the SSSI$^{(3)}$ order parameter, indicating that it is governed by structural interconversion between LDL-like and HDL-like local environments. Within a simple and effective two-state mapping, this structural evolution corresponds to an effective enthalpy scale of about 3$-$4 kJ mol$^{-1}$. 
{We notice that the heat of ice melting is about 6 kJ mol$^{-1}$, so that the effective enthalpy difference between LDL-like and HDL-like structures is nearly half of the enthalpy of ice melting.}
The two-state mapping yields a decomposition of $C_{P}$ into a physically constrained baseline and a structural contribution that captures both the supercooled maximum and the excess heat capacity at higher temperatures. 
The same qualitative trends are observed for both the NEP3@MB-pol and NEP3@revPBE0-D3 models (see section S9 \cite{supplemental}), indicating that the separation between a quantum-renormalized baseline and a structurally controlled anomalous contribution is robust across the two potentials considered. These results provide a microscopic framework, consistent with the LLPT scenario, for understanding how local structure controls the thermodynamic anomalies of water. 

\begin{acknowledgments}
This research was supported by the National Science Foundation under Grant No. 2305164.
This work used the Jetstream2 facility at Indiana University through allocation CHE220067 from the Advanced Cyberinfrastructure Coordination Ecosystem: 
Services and Support (ACCESS) program, which is supported by National Science Foundation grants \#2138259, \#2138286, \#2138307, \#2137603, and \#2138296.
\end{acknowledgments}

\section*{Supplemental Material}

The Supplemental Material includes: (1) details of the SSSI order parameter; (2) training details for the NEP3@MB-pol and NEP3@revPBE0-D3 potentials; (3) details of the classical MD and PIMD simulations; (4) results for the fit of the piecewise functions; (5) vDOS and cumulative $C_V$; (6) additional analysis of SSSI$^{(3)}$ and its correlation with enthalpy; (7) linearity of the residual background $C_{P, conf}$ term; (8) a sensitivity test for $\Delta H_{L/H}$ and $C_{P, conf}$; and (9) results for the NEP3@revPBE0-D3 potential.

\section*{Data Availability}
The data supporting the findings of this study, including the analysis tools, are available in the Zenodo archive doi:xxx. 

\section*{Conflict of Interest}
There are no conflicts to declare.

\nocite{*}

\bibliography{Water_Cp_SSSI}

\end{document}